\documentclass[%
 reprint,
 superscriptaddress,
 bibnotes,
 amsmath,amssymb,
 aps,
 prb,
 floatfix,
]{revtex4-2}

\usepackage{graphicx}
\usepackage{dcolumn}
\usepackage{bm}
\usepackage{hyperref}
\usepackage{float}
\usepackage{ulem}
\hypersetup{
  colorlinks   = true,
  urlcolor     = black,
  linkcolor    = black,
  citecolor   = blue
}
\usepackage{etoolbox}
\apptocmd{\sloppy}{\hbadness 10000\relax}{}{}
\usepackage{comment}

\usepackage{xcolor}

\newcommand{\trs}{$\mathcal{T}$}
\newcommand{\bM}{\mathbf{M}}

\begin{document}

\title{Circular Dichroism in Resonant Inelastic X-ray Scattering: Probing Altermagnetic Domains in MnTe}

\author{D. Takegami}
\affiliation{Department of Applied Physics, Waseda University, Shinjuku, Tokyo 169-8555, Japan}
\author{T. Aoyama}
\affiliation{Department of Physics, Graduate School of Science, Tohoku University, 6-3 Aramaki-Aoba, Aoba-ku, Sendai, Miyagi 980-8578, Japan}
\author{T. Okauchi}
\affiliation{Department of Physics and Electronics, Graduate School of Engineering,
Osaka Metropolitan University, 1-1 Gakuen-cho, Nakaku, Sakai, Osaka 599-8531, Japan}
\author{T. Yamaguchi}
\affiliation{Department of Physics and Electronics, Graduate School of Engineering,
Osaka Metropolitan University, 1-1 Gakuen-cho, Nakaku, Sakai, Osaka 599-8531, Japan}

\author{S. Tippireddy}
\affiliation{Diamond Light Source, Harwell Campus, Didcot OX11 0DE, United Kingdom}
\author{S. Agrestini}
\affiliation{Diamond Light Source, Harwell Campus, Didcot OX11 0DE, United Kingdom}
\author{M. Garc\'{i}a-Fern\'{a}ndez}
\affiliation{Diamond Light Source, Harwell Campus, Didcot OX11 0DE, United Kingdom}
\author{T. Mizokawa}
\affiliation{Department of Applied Physics, Waseda University, Shinjuku, Tokyo 169-8555, Japan}
\author{K. Ohgushi}
\affiliation{Department of Physics, Graduate School of Science, Tohoku University, 6-3 Aramaki-Aoba, Aoba-ku, Sendai, Miyagi 980-8578, Japan}
\author{Ke-Jin Zhou}
\affiliation{Diamond Light Source, Harwell Campus, Didcot OX11 0DE, United Kingdom}
\author{J. Chaloupka}
\affiliation{Department of Condensed Matter Physics, Faculty of Science, Masaryk University, Kotl\'{a}\v{r}sk\'{a}  2, 61137 Brno, Czech Republic}
\author{J. Kune\v{s}}
\email[]{kunes@physics.muni.cz}
\affiliation{Department of Condensed Matter Physics, Faculty of Science, Masaryk University, Kotl\'{a}\v{r}sk\'{a}  2, 61137 Brno, Czech Republic}
\affiliation{Institute for Solid State Physics, TU Wien, 1040 Vienna, Austria}
\author{A. Hariki}
\email[]{hariki@omu.ac.jp}
\affiliation{Department of Physics and Electronics, Graduate School of Engineering,
Osaka Metropolitan University, 1-1 Gakuen-cho, Nakaku, Sakai, Osaka 599-8531, Japan}
\author{H. Suzuki}
\email[]{hakuto.suzuki@tohoku.ac.jp}
\affiliation{Frontier Research Institute for Interdisciplinary Sciences, Tohoku University, Sendai 980-8578, Japan}
\affiliation{Institute of Multidisciplinary Research for Advanced Materials (IMRAM), Tohoku University, Sendai 980-8577, Japan}
\date{\today}

\begin{abstract}
X-ray magnetic circular dichroism provides a means to identify ferromagnetic, chiral, and altermagnetic orders via their time-reversal-symmetry (\trs) breaking. However, differentiating magnetic domains related by crystallographic symmetries remains a technical challenge. Here we reveal a circular dichroism (CD) in the resonant inelastic x-ray scattering (RIXS) spectra from the altermagnetic MnTe. The azimuthal dependence of the RIXS-CD intensity of the magnon excitations indicates a dominant occupation of a single altermagnetic domain. The RIXS-CD in our scattering geometry is ascribed to the mirror-symmetry breaking associated with the \trs-broken altermagnetic order. Our results establish RIXS-CD as a domain-sensitive probe of elementary excitations in quantum materials.
\end{abstract}

\maketitle

The interaction of circularly polarized photons with matter provides a fundamental probe of symmetry-breaking phenomena in solids. Left- and right-circularly polarized photons carry opposite helicities of spin angular momentum along their propagation direction. When they interact with a material, the corresponding transition amplitudes may differ depending on the spontaneously broken symmetries of the system. Circular dichroism (CD) in x-ray absorption has been extensively used to detect time-reversal symmetry (\trs) breaking in ferromagnets~\cite{Laan.G_etal.Phys.-Rev.-B1991,Chen.C_etal.Phys.-Rev.-Lett.1995} and, more recently, in some compensated magnets~\cite{Kimata.M_etal.Nat.-Commun.2021, Sakamoto.S_etal.Phys.-Rev.-B2021,Hariki.A_etal.Phys.-Rev.-Lett.2024}. In the dipole approximation, CD in absorption is strictly forbidden by \trs, meaning that a finite CD signal serves as a direct experimental signature of \trs-breaking in magnetic systems. Beyond time-reversal symmetry, circularly polarized photons are also sensitive to mirror symmetry operations, making CD a valuable tool for probing crystallographic symmetry. Recently, CD in inelastic photon scattering has been utilized to detect chiral phonons~\cite{Ishito.K_etal.Nat.-Phys.2023,Ueda.H_etal.Nature2023}, demonstrating its broader applicability in studying emergent symmetry-driven excitations.

Resonant inelastic x-ray scattering (RIXS) is a rapidly advancing technique that utilizes x-ray photons~\cite{Ament.L_etal.Rev.-Mod.-Phys.2011,Groot.F_etal.Nat.-Rev.-Methods-Primers2024,Mitrano.M_etal.Phys.-Rev.-X2024}. Unlike x-ray absorption, RIXS provides momentum resolution and is less constrained by spin conservation and dipole selection rules~\cite{Li.J_etal.Phys.-Rev.-X2023,Elnaggar.H_etal.Nat.-Commun.2023}. While the linear polarizations of the incoming light have been widely exploited to differentiate charge and spin excitations, the exploration of CD in RIXS (RIXS-CD) for magnetic materials remains largely uncharted. An unrecognized yet crucial property is the \textit{non-Hermiticity} of the RIXS process, discussed later, which inherently breaks \trs. Thus, RIXS-CD does not require $\mathcal{T}$-breaking, and non-zero RIXS-CD can exist even in the normal state. Yet, \trs-breaking affects the mirror symmetry properties of magnetic materials. With a careful choice of the polarizations and propagation directions of the two photons, RIXS-CD offers unique sensitivity to relativistic symmetry in magnetic systems, which we use here to study altermagnetism.

Altermagnets have emerged as a distinct subclass of collinear magnets, setting themselves apart from antiferromagnets~\cite{Smejkal.L_etal.Phys.-Rev.-X2022,Smejkal.L_etal.Phys.-Rev.-X2022_1,Smejkal.L_etal.Nat.-Rev.-Mater.2022}. Altermagnets are collinear magnets with zero net magnetization, where the antiferromagnetic sublattices are connected by proper or improper rotations rather than translation or inversion. Under these conditions, the magnetic ground states related by time-reversal operation become macroscopically inequivalent, giving rise to unique properties such as alternating spin polarization in electronic bands~\cite{Ahn.K_etal.Phys.-Rev.-B2019,Hayami.S_etal.J.-Phys.-Soc.-Jpn.2019}, the anomalous Hall effect~\cite{Gonzalez-Betancourt.R_etal.Phys.-Rev.-Lett.2023}, and spin currents~\cite{Naka.M_etal.Nat.-Commun.2019}.

\begin{figure*}[]
  \begin{center}
  \includegraphics[width=0.95\textwidth]{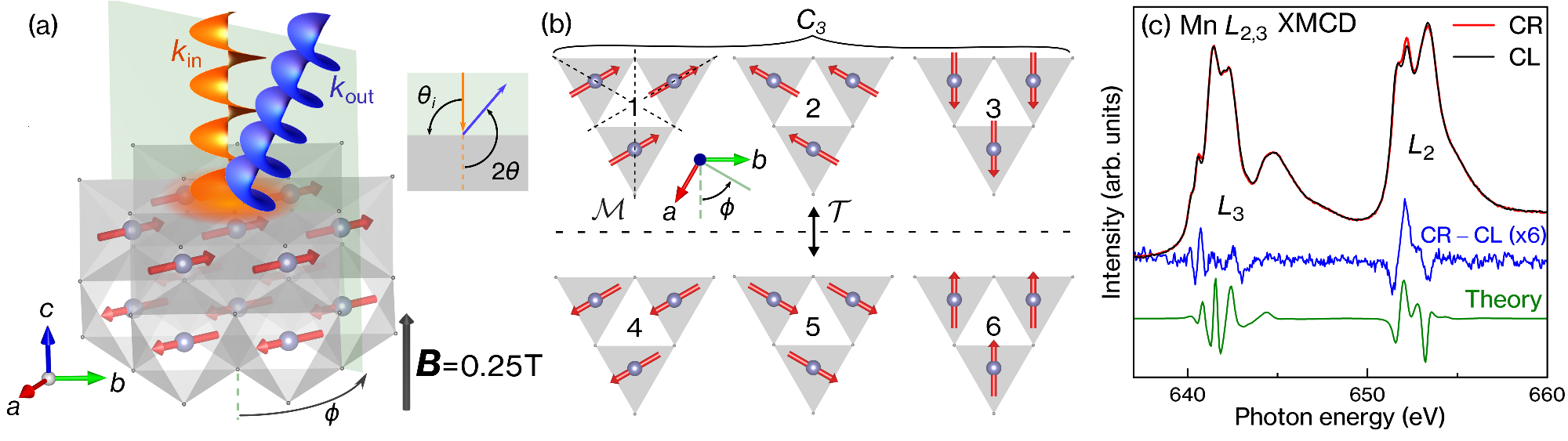}
  \end{center}
  \caption{(a) Altermagnetic order in MnTe and the scattering geometry of the resonant inelastic x-ray scattering (RIXS) experiment. The orange (blue) spiral depicts the incoming (outgoing) circularly polarized x-ray photons.
  $\theta_i$ denotes the angle between the incoming beam and the sample surface, $2\theta$ is the scattering angle, and the azimuthal angle $\phi$ generates the rotation of the scattering plane about the sample normal. 
  A small magnetic field of $B=$ 0.25 T is applied along the \textit{c} axis to stabilize the depicted altermagnetic domain.
  (b) Possible six altermagnetic domains. Domains 1-3 are \trs-equivalent domains related by the $C_3$ rotation, and domains 4-6 are their \trs-counterparts. The dotted lines on domain 1 indicate the mirror planes ($\mathcal{M}$) of the normal state.
  (c) Mn~$L_{2,3}$-edge x-ray absorption (XAS) spectra collected with total fluorescence yield (TFY) mode using the right (CR) and left (CL) circularly-polarized light. The blue curve shows the x-ray magnetic circular dichroism (XMCD) signal. The green curve shows the theoretical XMCD spectrum \cite{Hariki.A_etal.Phys.-Rev.-Lett.2024}. 
  }
  \label{Fig_Geom_XMCD}
  \end{figure*}

X-ray magnetic circular dichroism (XMCD) has proven effective for detecting \trs-breaking~\cite{Hariki.A_etal.Phys.-Rev.-Lett.2024} and determining the orientation of the N\'eel vector $\mathbf{L}=\mathbf{M}_1-\mathbf{M}_2$ in some altermagnets~\cite{Hariki.A_etal.Phys.-Rev.-B2024,Hariki.A_etal.2025}, where $\mathbf{M}_1$ and $\mathbf{M}_2$ are the sublattice magnetizations. However, in the archetypal altermagnet MnTe, where the two Mn ions are related by a screw operation [Fig.~\ref{Fig_Geom_XMCD}(a)], XMCD cannot distinguish three magnetic domains related by the $C_3$ rotation symmetry [Fig.~\ref{Fig_Geom_XMCD}(b)]. A way to overcome this difficulty is to combine XMCD with linear dichroism (LD) measurements \cite{Amin.O_etal.Nature2024}.
On the other hand, the scattering plane in RIXS-CD naturally distinguishes the $C_3$ domains from the spectral weight of the magnetic excitations.

In this Letter, we investigate Mn $L_{3}$-edge RIXS-CD in the altermagnetic state of MnTe through a combination of measurements and theoretical simulations. We observe a significant RIXS-CD signal in the magnon excitations, which aligns well with our numerical calculations. Interestingly, we find that RIXS-CD is not a consequence of \trs-breaking. Instead, it reflects the breaking of certain mirror symmetries induced by the \trs-broken altermagnetic order, enabling the distinction of magnetic domains in MnTe from the azimuthal angle dependence of the RIXS-CD signal.

{\it Experiment} --- Mn $L_3$-edge ($\sim$ 640 eV) RIXS experiment of MnTe single crystals was performed at the I21 beamline \cite{Zhou.K_etal.J.-Synchrotron-Rad.2022} of Diamond Light Source. Figure \ref{Fig_Geom_XMCD}(a) depicts the scattering geometry of the RIXS experiment. $\theta_i$ defines the angle between the incoming beam and the sample surface, and the scattering angle $2\theta$ was continuously varied to precisely define the momentum transfer ${\bm q}={\bm k}_{\rm in}-{\bm k}_{\rm out}$ in the three-dimensional reciprocal space. The azimuthal angle $\phi$ was used to change the in-plane ${\bm q}$ path. 
We applied a small magnetic field of $B=0.25$ T along the $c$ axis by mounting the single crystal onto a permanent magnet, to stabilize the equivalent \trs-domains [1-3 in Fig. \ref{Fig_Geom_XMCD}(b)] by utilizing the small canted magnetic moment along the $c$ axis \cite{Aoyama.T_etal.Phys.-Rev.-Mater.2024}. The total energy resolution of the spectrometer was set to 24 meV. The polarization of the outgoing photons was not analyzed. The momentum transfer ${\bm q}=(H, K, L)$ is expressed in the reciprocal lattice units (r.l.u.). The details of the characterization of the MnTe single crystals are provided in the Supplementary Material ~\cite{SM}.

{\it Theory} --- 
We simulate RIXS spectra using two complementary approaches: (i) an atomic calculation combined with the spin-wave theory of the Heisenberg model~\cite{Haverkort.M_etal.Phys.-Rev.-Lett.2010,Wang.R_etal.Phys.-Rev.-B2018}, and (ii) the DFT+DMFT Anderson impurity model (AIM) method~\cite{Hariki.A_etal.Phys.-Rev.-Lett.2018,Hariki.A_etal.Phys.-Rev.-B2020,Li.J_etal.Phys.-Rev.-X2023}.
The first approach captures the interference between excitations at different Mn sites, thereby accounting for the ${\bm q}$ dependence of the final states. However, it is limited to magnon excitations, and the time evolution of the intermediate state is constrained to a single atom.
The second method neglects the momentum dependence of the final states; the total intensity is treated as the sum of contributions from the individual Mn sites~\cite{Hariki.A_etal.Phys.-Rev.-Lett.2024,Winder.M_etal.Phys.-Rev.-B2020}. Nevertheless, it captures the evolution of the intermediate states and includes all types of excitations. In the first approach, the outgoing momentum determines the final state momentum and constrains the polarization of the outgoing photon. In the second approach, the outgoing momentum enters only through the photon polarization.

We first demonstrate a dominant population of the altermagnetic \trs-domains within the investigated x-ray spot by performing an XMCD measurement in the total fluorescent yield (TFY). Note that the XMCD and RIXS-CD measurements below are performed at the same spot. Figure~\ref{Fig_Geom_XMCD}(c) shows the x-ray absorption (XAS) spectra collected using the left (CL, black) and right (CR, red) circularly polarized photons. The XMCD signal (blue), defined as their difference, shows oscillatory behavior in the $L_3$ and $L_2$ regions. The XAS and XMCD signals near the $L_3$ edge are suppressed compared to those near the $L_2$ edge, because of a stronger self-absorption effect in fluorescence x-ray detection. 
However, the XMCD lineshape is in excellent agreement with the theoretical calculation (green), which exhibits oscillations characteristic of the altermagnetism in MnTe
and specific for the $[1\bar{1}00]$ orientation of the magnetic moments
~\cite{Hariki.A_etal.Phys.-Rev.-Lett.2024}. Note that this frequency dependence in the XMCD spectrum is distinct from the one originating from the net magnetization due to moment canting~\cite{Hariki.A_etal.Phys.-Rev.-Lett.2024}, confirming that the net magnetic moment remains tiny under a magnetic field of 0.25 T ~\cite{Kriegner.D_etal.Nat.-Commun.2016}. The XMCD amplitude at the $L_2$ edge closely matches the theoretical prediction for a single domain sample. This is in contrast to the previous XMCD spectrum of a MnTe thin film collected in the total electron yield \cite{Hariki.A_etal.Phys.-Rev.-Lett.2024}, whose amplitude was about ten times weaker than the theoretical prediction due to the coexistence of opposite \trs-domains. 
Note that the XMCD spectrum alone cannot distinguish the three domains connected by the $C_3$ rotation [domains 1-3 in Fig. \ref{Fig_Geom_XMCD}(b)]. RIXS-CD, on the other hand, allows the distinction of the $C_3$ domains, as we will show below.

\begin{figure}[]
  \begin{center}
  \includegraphics[width=1\columnwidth]{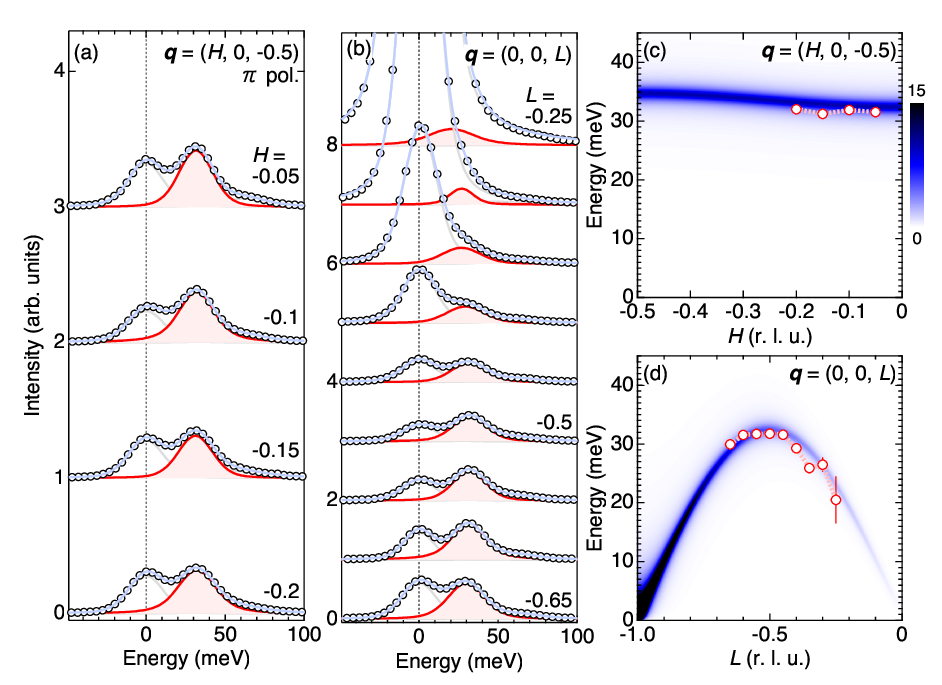}
  \end{center}
  \caption{(a), (b) Low-energy RIXS spectra of MnTe collected along the ${\bm q}=(H, 0, -0.5)$ and ${\bm q}=(0, 0, L)$ directions. The data were collected with $\pi$-polarized photons with $h\nu=$ 641.1 eV. Three Voigt profiles represent the elastic, magnon (red), and two-magnon excitations.  (c), (d) Transverse component of the spin dynamical structure factor $S({\bm q}, \omega)=(S^{yy}+S^{zz})({\bm q}, \omega)$. The experimental magnon peak energies are plotted as red circles.}
  \label{magnon}
  \end{figure}

Having confirmed the altermagnetic order, we now examine the magnon dispersion. Figure~\ref{magnon}(a) shows the low-energy region of the RIXS spectra along the ${\bm q} = (H, 0, -0.5)$ direction, collected with $\pi$-polarized $h\nu=$ 641.1 eV photons. One readily identifies the elastic peak, magnon excitation at $\sim$ 32 meV, and a shoulder structure of two-magnon excitations around $\sim$ 60 meV. We have fitted the spectra to three Voigt profiles, and the magnon contribution is highlighted with shaded red curves. In agreement with the former INS results in Refs.~\cite{Szuszkiewicz.W_etal.Phys.-Rev.-B2006,Liu.Z_etal.Phys.-Rev.-Lett.2024}, the magnon energy shows a weak dependence on the in-plane $H$, reflecting subdominant in-plane exchange interactions.
On the other hand, the magnon exhibits significant dispersion along the ${\bm q} = (0, 0, L)$ direction [Figure~\ref{magnon}(b)], due to the dominant out-of-plane exchange interactions. The magnon energy is maximal at $L=-0.5$ and shows a downward turn as $L$ approaches $0$ or $-1$.

To describe the magnon dispersion, we use the following $S=5/2$ spin Hamiltonian:
\begin{equation}
\mathcal{H} = \sum_{\langle i,j \rangle} J_{ij} \mathbf{S}_i \cdot \mathbf{S}_j + D \sum_i (S_i^z)^2, \notag
\end{equation}
where $J_{ij}$ represents the Heisenberg interactions between spins on the $i$-th and $j$-th Mn sites, and $D$ is the single-ion anisotropy term. Figures~\ref{magnon}(c) and (d) compare the experimental magnon dispersions (red circles) with the transverse component of the spin dynamical structure factor, $S({\bm q}, \omega)=(S^{yy}+S^{zz})({\bm q}, \omega)$, where the $x$ axis is taken parallel to the N\'eel vector $\mathbf{L}$. The optimal parameters, computational methods, and high-energy multiplets that confirm the $S=5/2$ ground states are detailed in the Supplemental Material~\cite{SM}. $S({\bm q}, \omega)$ along ${\bm q} = (H, 0, -0.5)$ in Fig.~\ref{magnon}(c) exhibits an almost flat $H$ dependence, reproducing the experimental data within the accessible ${\bm q}$ space. Also, $S({\bm q}, \omega)$ along ${\bm q} = (0, 0, L)$ in Fig.~\ref{magnon}(d) reaches a maximum at $L = -0.5$ and approaches $\omega=0$ as $L$ approaches $0$ or $-1$, in agreement with the RIXS data and the INS results \cite{Liu.Z_etal.Phys.-Rev.-Lett.2024}. The splitting of the two magnon dispersions along these high-symmetry ${\bm q}$ paths is almost negligible except near the bottom of the band, as it is solely caused by the relativistic $D$ term.

\begin{figure}[]
  \begin{center}
  \includegraphics[width=1\columnwidth]{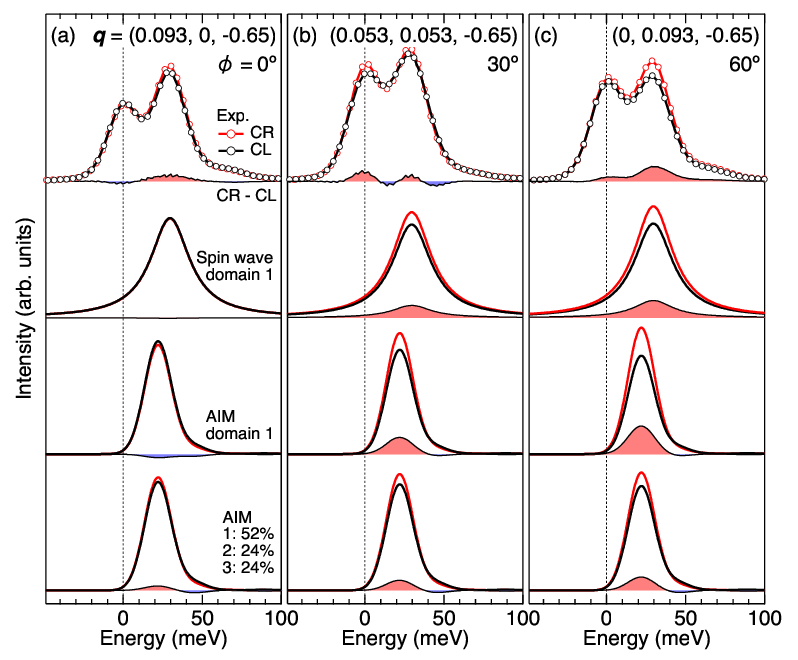}
  \end{center}
  \caption{(a)-(c) RIXS spectra collected with right and left circularly polarized light, at the azimuthal angles $\phi=0^\circ$, $30^\circ$, and $60^\circ$, respectively. RIXS-CD intensity defined as their difference is also shown. The curves in the second and third rows are the corresponding theoretical RIXS and RIXS-CD spectra for the domain 1 computed with the spin wave and Anderson impurity model (AIM) methods. The bottom curves show the fit of the experimental curves as superpositions of the theoretical AIM spectra for the domains 1-3.}
  \label{rixsmcd}
  \end{figure}

Next, we investigate the RIXS-CD at magnon excitations.
As depicted in Fig. \ref{Fig_Geom_XMCD}(a), circularly-polarized x-ray photons were incident parallel to the $c$ axis ($\theta_{i}=90^\circ$), and the outgoing photons were detected at the scattering angle of $2\theta = 150^\circ$. The top panels in Figs.~\ref{rixsmcd}(a)-(c) display the RIXS spectra collected with circularly polarized photons, obtained at three azimuthal angles $\phi = 0^\circ$, $30^\circ$, and $60^\circ$, respectively. For the real-space definition of $\phi = 0^\circ$, see Fig.~\ref{Fig_Geom_XMCD}(b). The RIXS-CD spectra are also presented, with positive and negative regions shaded in red and blue, respectively. The momentum transfers are ${\bm q} = (0.093, 0, -0.65)$, $(0.053, 0.053, -0.65)$, and $(0, 0.093, -0.65)$, all sharing the same out-of-plane component $L = -0.65$ but differing in their projections to the $ab$-plane. 
The CD at the magnon peak is clearly observed at $\phi=60^\circ$, while the CD at $\phi=0^\circ$ and $30^\circ$ is weaker.

To understand this $\phi$ dependence in RIXS-CD, the curves in the second and third rows in Fig.~\ref{rixsmcd} show the theoretical RIXS and RIXS-CD spectra obtained using the spin wave and AIM methods, respectively. The calculations are performed for the single domain 1 in Fig.~\ref{Fig_Geom_XMCD}(b). 
First, we find that both methods yield maximal CD intensities at $\phi=60^\circ$, and negligibly small CD at $\phi=0^\circ$. Note that these two directions correspond to the largest amplitude of the $g$-wave altermagnetism of MnTe with the opposite signs. Correspondingly, the two magnon bands with opposite chiralities show energy splitting in opposite directions. Nevertheless, the RIXS-CD intensity does not show a simple sign reversal with the same amplitude, both in the experiment and theory. This demonstrates that the RIXS-CD amplitude does not reflect the altermagnetic structure of magnons in the ${\bm q}$ space. Thus, RIXS-CD is not a direct probe of the chirality of magnons, as claimed in Ref. \cite{Jost.D_etal.2025}. Instead, the RIXS-CD is governed by the orientation of the outgoing photon polarization with respect to the $\mathbf{L}$ vector of the ground state. This allows us to use RIXS-CD to distinguish the $C_3$ domains.

Here, we clarify the origin of CD in the RIXS process. In general, the Kramers-Heisenberg formula for the RIXS cross section $F^\pm(\bm{k}_\text{out})\equiv\frac{\delta^2\sigma_\pm}{\delta \Omega \delta \omega}(\bm{k}_\text{out})$~\cite{Kramers.H_etal.Z.-Phys.1925},
\begin{eqnarray*}
F^\pm(\bm{k}_\text{out})
\propto & \sum\limits_{f,\alpha} \Bigg| \sum\limits_{m} \dfrac{ \langle f | \operatorname{V}_{\alpha} |m\rangle \langle m | 
\operatorname{V}_{\pm}| i \rangle}{\omega_{\rm in} + E_i - E_m + i\Gamma} \Bigg|^2 \notag \\ 
& \times \delta(\omega_{\rm in} - \omega_{\rm out} + E_i - E_f)\\
\equiv & \sum\limits_{f,\alpha} \left| \sum\limits_{m} \langle f | \operatorname{V}_{\alpha} \operatorname{X}
\operatorname{V}_{\pm}| i \rangle \right|^2  \delta(\ldots),
\end{eqnarray*}
has the form of the Fermi golden rule connecting the initial $|i\rangle$ and final $|f\rangle$ states with respective energies $E_i + \omega_{\rm in}$ and $E_f + \omega_{\rm out}$. 
Notably, the operator
$\operatorname{X}= \sum\limits_{m} \dfrac{   |m\rangle \langle m |  }{\omega_{\rm in} + E_i - E_m + i\Gamma}$, which describes the propagation in the intermediate state, 
is non-Hermitian due to the imaginary part of the denominator. The non-Hermiticity arises from the resonant processes, where the initial and final states (including photons) are degenerate with a continuum of intermediate states. This property persists even as $\Gamma \to 0$. Consequently, unlike XMCD, RIXS is not inherently sensitive to $\mathcal{T}$-breaking, and non-zero CD can exist even in the normal state.
However, unitary symmetries, such as crystallographic symmetries, can impose selection rules that suppress RIXS-CD. Since the incoming and outgoing photons are part of the initial and final states ($|i\rangle$ and $|f\rangle$), their momenta ($\bm{k}_{\rm in}$ and $\bm{k}_{\rm out}$) must be included in the symmetry analysis. For instance, RIXS-CD is forbidden if both $\bm{k}_{\rm in}$ and $\bm{k}_{\rm out}$ lie in the same mirror plane. In the present geometry, this symmetry argument leads to the vanishing of RIXS-CD at $\phi = (n\times 60)^\circ$ in the normal state [see the mirror planes ($\mathcal{M}$) in Fib. \ref{Fig_Geom_XMCD}(b)]. Thus, the non-zero RIXS-CD at $\phi=0^\circ$ and $60^\circ$ in our experiment is allowed by the mirror-symmetry breaking associated with the altermagnetic order. 

In the AIM approximation, the selection rule becomes even more restrictive. 
The total scattered intensity averaged over outgoing light polarizations is then given by~\cite{SM}
\begin{eqnarray}
\label{eq:tensor}
   F^\pm(\bm{k}_\text{out})\!\propto\!
   \sum_f\left(\operatorname{Tr}(\bM^\pm_{fi}\overline{\bM}^\pm_{fi})- \hat{\mathbf{k}}\!\cdot\!\bM^\pm_{fi}\overline{\bM}^\pm_{fi}\!\cdot\!\hat{\mathbf{k}}\right),\notag
\end{eqnarray}
with $\bM^\pm_{fi}=\langle f|\mathbf{V}\operatorname{X}\operatorname{V}^\pm|i\rangle$ and $\hat{\mathbf{k}}=\bm{k}_\text{out}/|\bm{k}_\text{out}|$.
This implies the $F^\pm_n(\phi)=A^{\pm}+B^\pm\cos(2\phi+\tfrac{2}{3}\pi n+ \phi^\pm_0)$ dependence of the RIXS amplitude, where $n=1,2,3$ is the index of the $C_3$ domain in Fig. \ref{Fig_Geom_XMCD}(b). As a result, an equal domain population leads to a $\phi$-constant RIXS-CD signal ($\Delta F_n=F_n^{+}-F_n^{-}$). The mirror-plane relationship between domains
1 and 4 implies $\Delta F_4(\phi)=-\Delta F_1(-\phi-\tfrac{2\pi}{3})$. The \trs-domains thus lead to a RIXS-CD signal with opposite constant part and mirror
$\phi$-dependence. Since the mirror plane is a symmetry of the normal state, the RIXS-CD vanishes for arbitrary $\phi$ in the present geometry. Nevertheless, for a general $\bm{k}_{\rm in}$ CD is allowed even in the normal state. For details see the Supplemental Material~\cite{SM}.

The strong azimuthal dependence of the experimental RIXS-CD, therefore, points to a considerable difference in the domain populations within the measurement spot. We fitted the experimental RIXS-CD curves to superpositions of the AIM curves of the three domains [1-3 in Fig.~\ref{Fig_Geom_XMCD}(b)]. Here, the population ratio was optimized to best reproduce the angular dependence of the magnon RIXS-CD spectral weight. The fitted curves, shown at the bottom of Figs.~\ref{rixsmcd}(a-c), show good agreement with the experimental data for all the three $\phi$ angles, including the overall positive sign originating from the constant term of $F^\pm_n(\phi)$. As expected from the largest CD at $\phi = 60^\circ$, domain 1 has the dominant occupation of 52~\%, while domains 2 and 3 also have non-negligible occupations of 24~\%, and 24~\%, respectively. RIXS-CD thus allows the estimation of magnetic domain weights, from the spectral weight of the magnon excitations.

In conclusion, we have conducted an Mn $L_3$ RIXS study of the magnetic excitations in the altermagnetic MnTe. The azimuthal dependence of the RIXS-CD intensity reveals a dominant occupation of a single altermagnetic domain. The observed non-zero RIXS-CD in our scattering geometry is attributed to the mirror-symmetry breaking associated with the \trs-breaking altermagnetic order. Our findings establish RIXS-CD as a domain-sensitive probe of elementary excitations in a broad range of transition metal compounds. RIXS-CD detects the unitary (space) symmetry breaking via \trs-breaking magnetic orders. This makes RIXS-CD a complementary technique to XMCD, with potential applications to a wider class of anomalous Hall antiferromagnets, including the non-collinear ones \cite{Nakatsuji.S_etal.Annu.-Rev.-Condens.-Matter-Phys.2022,Cheong.S_etal.npj-Quantum-Mater.2024}. 

We thank Y. Motome and S. Souma for enlightening discussions. This work was supported by JSPS KAKENHI Grant Numbers JP22K13994, JP21K13884, JP23K03324, JP23H03817. J.C. and J.K. acknowledge financial support by Quantum Materials for Applications in Sustainable Technologies Grant No. CZ.02.01.01/00/22\_008/0004572.
D.T. acknowledges the financial support by the Deutsche Forschungsgemeinschaft (DFG, German Research Foundation) under the Walter Benjamin Programme, Projektnummer 521584902. Part of the computations in this work were performed using the facilities of the Supercomputer Center, the Institute for Solid State Physics, the University of Tokyo.

\bibliography{MnTe}

\end{document}


\title{Supplemental Material for\\ Circular Dichroism in Resonant Inelastic X-ray Scattering: Probing Altermagnetic Domains in MnTe}

\author{D. Takegami}
\affiliation{Department of Applied Physics, Waseda University, Shinjuku, Tokyo 169-8555, Japan}
\author{T. Aoyama}
\affiliation{Department of Physics, Graduate School of Science, Tohoku University, 6-3 Aramaki-Aoba, Aoba-ku, Sendai, Miyagi 980-8578, Japan}
\author{T. Okauchi}
\affiliation{Department of Physics and Electronics, Graduate School of Engineering,
Osaka Metropolitan University, 1-1 Gakuen-cho, Nakaku, Sakai, Osaka 599-8531, Japan}
\author{T. Yamaguchi}
\affiliation{Department of Physics and Electronics, Graduate School of Engineering,
Osaka Metropolitan University, 1-1 Gakuen-cho, Nakaku, Sakai, Osaka 599-8531, Japan}

\author{S. Tippireddy}
\affiliation{Diamond Light Source, Harwell Campus, Didcot OX11 0DE, United Kingdom}
\author{S. Agrestini}
\affiliation{Diamond Light Source, Harwell Campus, Didcot OX11 0DE, United Kingdom}
\author{M. Garc\'{i}a-Fern\'{a}ndez}
\affiliation{Diamond Light Source, Harwell Campus, Didcot OX11 0DE, United Kingdom}
\author{T. Mizokawa}
\affiliation{Department of Applied Physics, Waseda University, Shinjuku, Tokyo 169-8555, Japan}
\author{K. Ohgushi}
\affiliation{Department of Physics, Graduate School of Science, Tohoku University, 6-3 Aramaki-Aoba, Aoba-ku, Sendai, Miyagi 980-8578, Japan}
\author{Ke-Jin Zhou}
\affiliation{Diamond Light Source, Harwell Campus, Didcot OX11 0DE, United Kingdom}
\author{J. Chaloupka}
\affiliation{Department of Condensed Matter Physics, Faculty of Science, Masaryk University, Kotl\'{a}\v{r}sk\'{a}  2, 61137 Brno, Czech Republic}
\affiliation{Central European Institute of Technology, Masaryk University, Kamenice 753/5, 62500 Brno, Czech Republic}
\author{J. Kune\v{s}}
\affiliation{Department of Condensed Matter Physics, Faculty of Science, Masaryk University, Kotl\'{a}\v{r}sk\'{a}  2, 61137 Brno, Czech Republic}
\affiliation{Institute for Solid State Physics, TU Wien, 1040 Vienna, Austria}
\author{A. Hariki}
\affiliation{Department of Physics and Electronics, Graduate School of Engineering,
Osaka Metropolitan University, 1-1 Gakuen-cho, Nakaku, Sakai, Osaka 599-8531, Japan}
\author{H. Suzuki}
\affiliation{Frontier Research Institute for Interdisciplinary Sciences, Tohoku University, Sendai 980-8578, Japan}
\affiliation{Institute of Multidisciplinary Research for Advanced Materials (IMRAM), Tohoku University, Sendai 980-8577, Japan}

\maketitle

\section{\NoCaseChange{Characterization of MnTe single crystals}}
\begin{figure}[ht]
\begin{center}
\includegraphics[width=5cm]{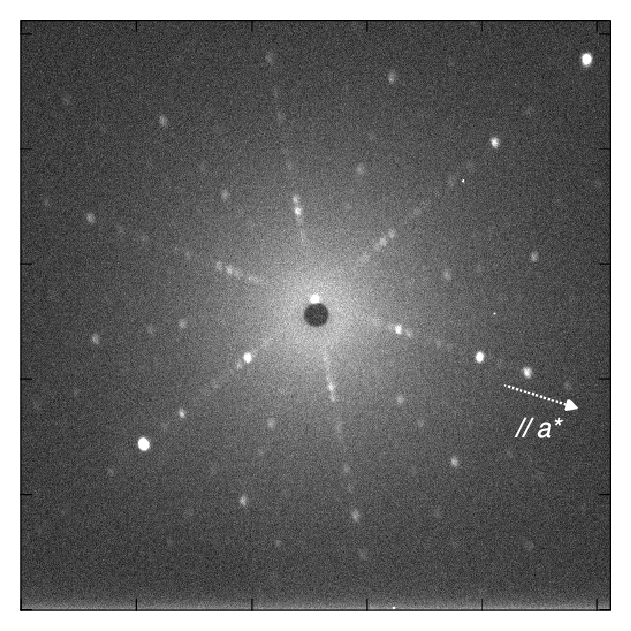}
\end{center}
\caption{Laue diffraction pattern of the MnTe single crystal measured in the RIXS experiment. }
\label{Laue}
\end{figure}

The single crystals measured in the RIXS experiment were grown by the procedure outlined in Ref. \cite{Aoyama.T_etal.Phys.-Rev.-Mater.2024}. The orientation of the crystal lattice was determined with Laue diffraction. The obtained pattern is shown in Fig. \ref{Laue}. Single crystals from the same batch were previously used in angle-resolved photoemission measurements  \cite{Osumi.T_etal.Phys.-Rev.-B2024}, which confirmed their high crystallinity and the altermagnetic spin degeneracy lifting of the electronic bands.

\section{\NoCaseChange{Incident energy dependence of RIXS signal and the multiplet structures}}
\begin{figure*}[ht]
\begin{center}
\includegraphics[width=12cm]{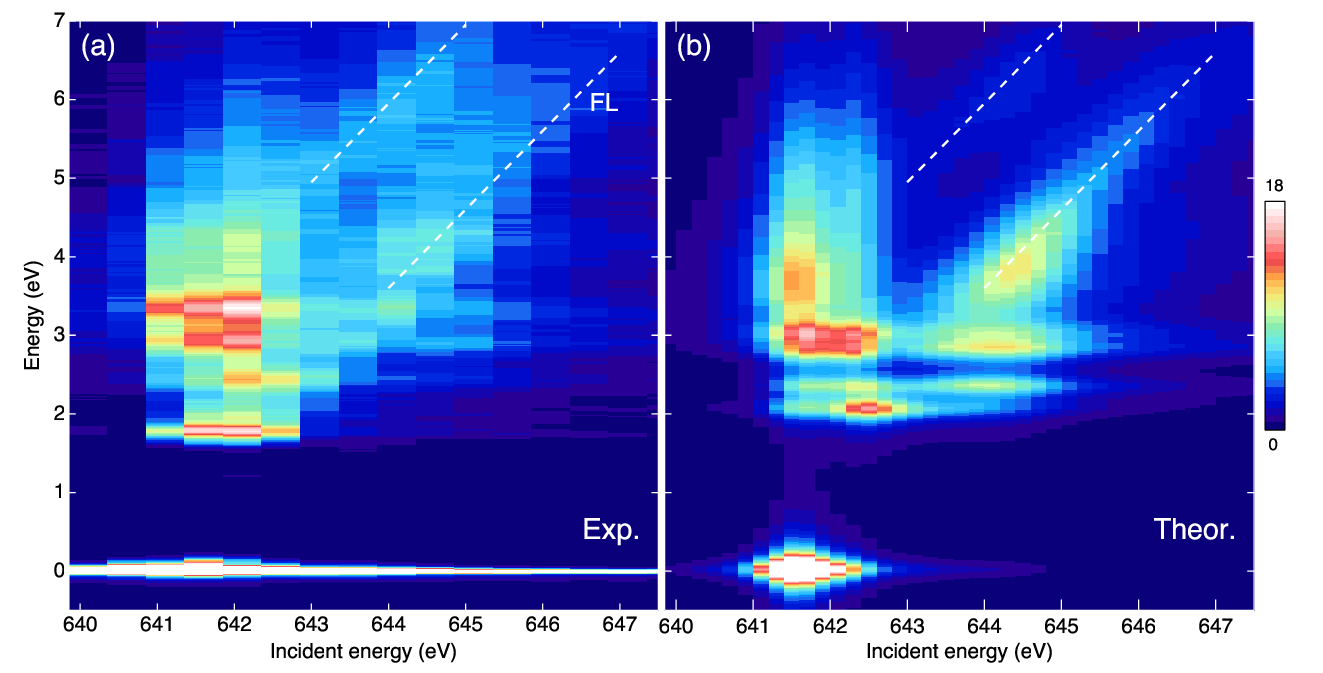}
\end{center}
\caption{(a) Colormap of the incident energy dependence of the high-energy RIXS spectra. Dotted lines indicate the fluorescence-like (FL) signal. (b) Colormap of theoretical RIXS spectra computed with the LDA+DMFT method ~\cite{Hariki.A_etal.Phys.-Rev.-B2020}.} 
\label{hvmap}
\end{figure*}

Figure \ref{hvmap}(a) shows the colormap of the high-energy RIXS spectra collected with incident x-ray photons around the Mn $L_3$ absorption edge ($640.1\leq h\nu\leq 648.1$ eV). The RIXS data were collected in a medium-resolution optical setup with a total energy resolution of 32 meV in FHWM. The main Raman-like features, which have constant energy loss as a function of $h\nu$, represent the local multiplet excitations within the Mn$^{2+}$ ions. In addition, fluorescent-like (FL) signals (dotted lines) appear in the region $h\nu \gtrsim$ 642 eV, whose energy loss evolves linearly with the incident energy (dotted lines). The FL signals represent the excitations to the final states with itinerant character. 

We reproduce the incident energy dependence of the RIXS signal using the local density approximation (LDA) + dynamical mean-field theory (DMFT) method \cite{Hariki.A_etal.Phys.-Rev.-B2020}. Figure \ref{hvmap}(b) shows the colormap of the theoretical RIXS intensity. It captures the Raman-like features and the FL signals, with correct intensity. The Raman-like signals are attributed to the $dd$ excitations from the high-spin $S=5/2$ ground state to the higher-energy multiplets. The theoretical RIXS spectrum also reproduces the linear dependence of energy loss of the FL signals with the incident energy.


\section{\NoCaseChange{Computation of spin susceptibility}}

Here, we describe the computation of spin susceptibility in the altermagnetic state of MnTe. The divalent Mn ions with the localized $3d^5$ electrons host the spin $S = 5/2$ in the high-spin ground state. To describe the low-energy magnetic excitations, we have employed the following $S = 5/2$ spin Hamiltonian:
\begin{equation}
\mathcal{H} = \sum_{\langle i,j \rangle} J_{ij}  \mathbf{S}_i \cdot \mathbf{S}_j + D \sum_i (S_i^z)^2, \notag
\end{equation}
where $J_{ij}$ represents the exchange interactions between the spins on the $i$-th and $j$-th Mn sites, and $D$ is the single-ion anisotropy
constant. We consider the exchange interactions $J_i$ up to the third nearest-neighbor Mn-Mn bonds. 
We employ the parameter values $J_1 = 3.505$~meV, $J_2 = -0.115$~meV, $J_3 = 0.495$~meV, and $D = 0.022$~meV, following the inelastic neutron scattering study~\cite{Szuszkiewicz.W_etal.Phys.-Rev.-B2006}. In this work, a small adjustment of the dominant $J_1$ term was made to reproduce the RIXS data more accurately.
We then transform the spin model to the Holstein-Primakoff boson representation followed by a standard linearized spin-wave treatment, assuming the antiferromagnetic spin alignment with the N\'{e}el vector of ${\bf L}=[1\overline{1}00]$. 
Subsequently, the transverse component of the spin dynamical structure factor, $S({\bm q}, \omega)=(S^{yy}+S^{zz})({\bm q}, \omega)$, was calculated for the spin-wave Hamiltonian. 

\section{\NoCaseChange{Unitary and anti-unitary transformations of RIXS cross section}}
Next, we consider the effect of a unitary symmetry in the RIXS cross section. If the system is transformed by
unitary operation $\mU$
\begin{equation}
    \begin{split}
    \tilde{F}^\pm(\bk)&=\sum_{f,\alpha}\left|\sum_m\frac{\expval{\mU f|V_{\alpha}|\mU m}
    \expval{\mU m|V_{\pm}|\mU i}}{E_i-E_m+i\Gamma}\right|^2 \delta(\ldots)\\
     &= \sum_{f,\alpha}\left|\sum_m\frac{\expval{ f|\mU^{-1}V_{\alpha}\mU |m}\expval{ m|\mU^{-1}V_{\pm}\mU| i}}
    {E_i-E_m+i\Gamma}\right|^2 \delta(\ldots) \\
    &=F^s(\bk').
    \end{split}
\end{equation}
In the second line, we use the unitarity of $\mU$ to transform the operators instead of functions. Note, that $\mU$
may or may not be the symmetry of the system. In the former case, this equation constrains $F^\pm(\bk)$. In the latter case, e.g., representing transformation between different domains, the equation relates the $F^\pm(\bk)$ in these
domains. The dipole operators transform as vectors or pseudovectors. In the present geometry rotation by an angle
$\gamma$ around the $c$-axis lead the $V^\pm$ operator unchanged, while the outgoing operators rotate as vectors leading to
\begin{equation*}
    \tilde{F}^\pm(\phi)=F^{\pm}(\phi+\gamma).
\end{equation*}
A mirror plane containing the $c$ axis switches the circular polarization of the incoming operator, while the outgoing operator again transforms as a vector leading to
\begin{equation*}
    \tilde{F}^\pm(\phi)=F^{\mp}(-\phi),
\end{equation*}
where we have assumed that the angle $\phi$ is measured from the mirror plane.

The anti-unitary operations, such as time reversal $\mT$, act on the RIXS intensity as follows
\begin{equation}
    \begin{split}
    \tilde{F}^\pm(\bk)&=\sum_{f,\alpha}\left|\sum_m\frac{\expval{\mT f|V_{\alpha}|\mT m}
    \expval{\mT m|V_{\pm}|\mT i}}{E_i-E_m+i\Gamma}\right|^2 \delta(\ldots)\\
     &= \sum_{f,\alpha}\left|\sum_m\frac{\overline{\expval{ f|\mT^{-1}V_{\alpha}\mT |m}}\overline{\expval{ m|\mT^{-1}V_{\pm}\mT| i}}}
    {E_i-E_m+i\Gamma}\right|^2 \delta(\ldots). 
    \end{split}
\end{equation}
Since both the denominator and numerator are in general complex numbers, there is no transformation connecting $\tilde{F}$ and $F$, calculated with $\mT$-related wave functions. The root cause is the presence
of the imaginary part $i\Gamma$ in the denominator, which makes the evolution operator in the intermediate state
\begin{equation*}
    \operatorname{X}= \sum\limits_{m} \dfrac{  |m\rangle \langle m | 
}{\omega_{\rm in} + E_i - E_m + i\Gamma}
\end{equation*}
non-Hermitian. While the common interpretation of $\Gamma$ is a finite core-hole lifetime, the non-Hermiticity does not go away even in the limit of an infinitely lived core hole. In this case, one has to take the limit
$\Gamma\rightarrow 0^+$, which generates non-Hermiticity as long as there is a finite density of intermediate states
$\sim\sum_m \delta(E_i-E_m)$ degenerate with the initial state (including the photon). 
The non-Hermiticity therefore originates in the irreversible nature of the decay of the initial state into
the continuum of the intermediate states.

\section{\NoCaseChange{RIXS-CD amplitude in the impurity approximation}}
To derive the angular distribution of RIXS intensity, we define a transition operator
\begin{equation*}
   \begin{split}
           [\operatorname{\bM^\pm]_\alpha}\equiv \operatorname{V}_{\alpha} \operatorname{X}
\operatorname{V}_{\pm}\\
           \bM^\pm_{fi}\equiv\expval{f|\bM^\pm|i}
   \end{split}
\end{equation*}
The total scattered intensity can be obtained as a sum of intensities over two linear polarizations $\hat{\mathbf{e}}_\alpha$
perpendicular to $\bm{k}_\text{out}$. 
\begin{equation*}
    \begin{split}
      F^\pm(\bm{k}_\text{out})&\propto \sum_f\sum_\alpha |\hat{\mathbf{e}}_\alpha\cdot \bM^\pm_{fi}|^2\delta(\ldots) \\
           &= \sum_f\left(|\bM^\pm_{fi}|^2-|\hat{\bk}\cdot\bM^\pm_{fi}|^2\right) \delta(\ldots)\\
           &= \sum_f\left(\operatorname{Tr}(\bM^\pm_{fi}\overline{\bM}^\pm_{fi})-
           \hat{\bk}\cdot\bM^\pm_{fi}\overline{\bM}^\pm_{fi}\cdot\hat{\bk} \right)\delta(\ldots),\\
    \end{split}
\end{equation*}
where $\hat{\bk}=\bm{k}_\text{out}/|\bm{k}_\text{out}|$.
This formula has a clear interpretation as the total amplitude for any polarization minus the contribution of the unphysical longitudinal polarization. Since the $\bM\overline{\bM}$ tensor does not depend on $\bm{k}_\text{out}$ in the
impurity approximation, the dependence of RIXS as well as RIXS-CD on $\phi$ is fully captured by the second term.

\begin{figure}[t]
\begin{center}
\includegraphics[width=1\columnwidth]{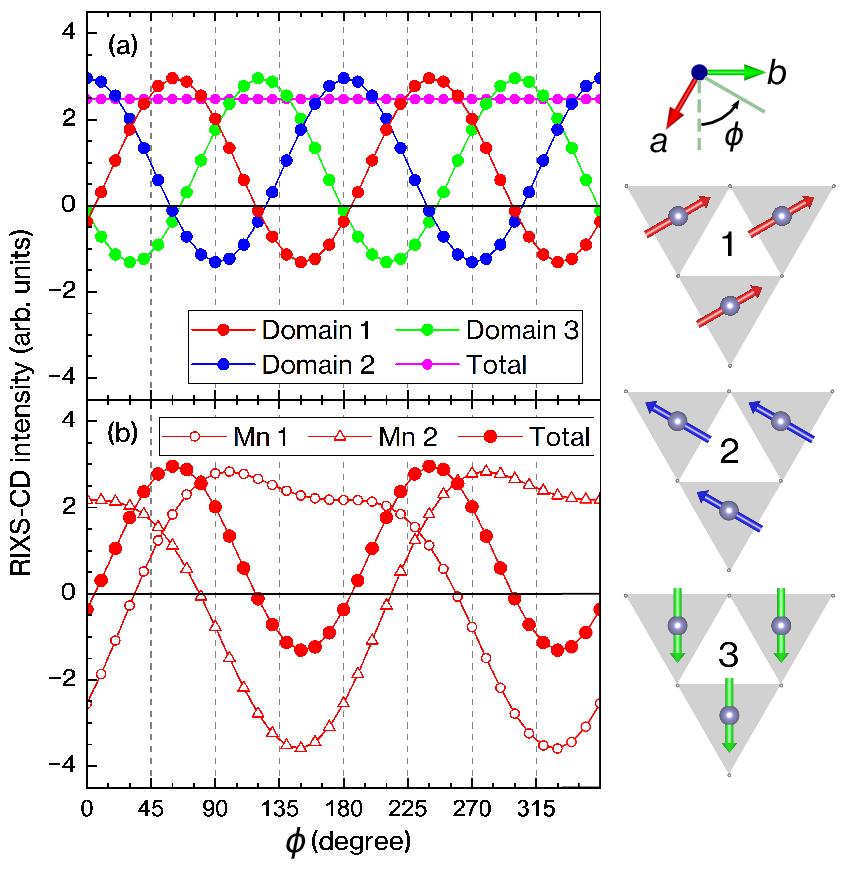}
\end{center}
\caption{(a) Azimuthal angle ($\phi$) dependence of the RIXS-CD intensity in the single-magnon excitation computed by the LDA+DMFT AIM method, for the three $C_3$ domains shown in the right. The equal sum of the contributions from the three domains results in a $\phi$-independent CD, as indicated by the pink curve. (b) Mn site-resolved contributions to the RIXS-CD in domain 1.}
\label{mcd_phidep}
\end{figure}

The single-site contribution is thus a sum of terms proportional to 1, $\cos(\phi)$ and $\cos(2\phi)$. Thanks to the 
altermagnetic order with the two sublattice related by $C_2$ rotation the term proportional to $\cos(\phi)$ vanishes
upon summation over the magnetic sublattices, see Fig.~\ref{mcd_phidep}. Since $\bm{k}_\text{in}\parallel c$
is invariant under the $C_3$ rotations, the $\phi$-distributions of the RIXS intensities corresponding to the $C_3$-domains 
are rotated by $\tfrac{2}{3}\pi$:
\begin{equation*}
    F^\pm_n(\phi)=A^\pm + B^\pm\cos(2\phi+\phi^\pm_0+\tfrac{2}{3}\pi n).
\end{equation*}
The parameters $A^\pm$, $B^\pm$ and $\phi^\pm_0$ depend the matrix elements $\bM$ and polar angle $\Theta$.

Next, we consider the time-reversed domains with N\'eel vectors $\mathbf{L}$ and $-\mathbf{L}$. For the easy-axis direction
$\mathbf{L}$ belongs to a mirror plane of the crystallographic space group and the states with $\mathbf{L}$ and $-\mathbf{L}$
are related by this mirror operation. Let us consider domains 1 and 4 the mirror plane for which corresponds to $\phi=-\tfrac{\pi}{3}$.
Under this operation the orientation of $\bm{k}_\text{out}$ changes from $\phi$ to $-\phi-\tfrac{2\pi}{3}$ and the polarization
of the incoming light changes from right- to left-circularly polarized and vice versa. This leads to
\begin{equation*}
    F^\pm_4(\phi)=F^\mp_1(-\phi-\tfrac{2\pi}{3}).
\end{equation*}
Similar relations apply for the other $\mathbf{L}$ and $-\mathbf{L}$ pairs.

\section{\NoCaseChange{RIXS cross section on the magnon dispersion}}

We describe the approach (i) in the main text for simulating the RIXS cross sections. This approach combines an atomic calculation for the RIXS excitation amplitudes with the spin-wave theory for the spin model described in Sec.~III. Our implementation follows Refs.~\cite{Haverkort.M_etal.Phys.-Rev.-Lett.2010,Wang.R_etal.Phys.-Rev.-B2018}. The RIXS cross section with circularly polarized incident x-rays is given by the Kramers-Heisenberg formula,
\begin{equation}
\begin{split}
	F^{\pm}
    &\varpropto
	\sum_{f,\alpha} \Big{|} \sum_m \frac{ \langle f | V_{\alpha} |m\rangle \langle m | V_{\pm}| i \rangle}{\omega_{\rm in}+E_i-E_m+i\Gamma} \Big{|} ^{2}  
	\delta(\omega_{\rm loss}+E_i-E_f)  \\
	\vspace{+0.3cm}
	&=
	-\frac{1}{\pi} {\rm Im} \ \langle i | R^{\dag}_{\textit{\textbf{q}}, \omega_{\rm in}}
	\frac{1}{\omega_{\rm loss}+E_i-H+i\delta}  
	R_{\textit{\textbf{q}}, \omega_{\rm in}}| i \rangle, \notag 
		\label{rixs}
\end{split}
\end{equation}
where $H$ is the Hamiltonian of the whole system.
The operator $R_{\bm{q}, \omega_{\rm in}}$ is defined as 
\begin{eqnarray}
	R_{\textit{\textbf{q}}, \omega_{\rm in}}&=&
	V_{\alpha} \frac{1}{\omega_{\rm in}+E_i-H+i\Gamma} V_{\pm}  \notag \\
	&=&
	\sum_{j} e^{-i \bm{q} \cdot \bm{r}_{j}}
	V_{\alpha}( \bm{r}_j ) \frac{1}{\omega_{\rm in}+E_i-H+i\Gamma} V_{\pm}( \bm{r}_j  ), \notag     
\end{eqnarray}
where $V$ are expanded at each Mn atom at ${\bm r}_j$ as
$V_{\pm}=\sum_{j} V_{\pm}( {\bm r}_j ) e^{i\bm{k}_{\rm in } \cdot \bm{r}_{j} }$ and
$V_{\alpha}=\sum_{j} V_{\alpha}( {\bm r}_j ) e^{i\bm{k}_{\rm out} \cdot \bm{r}_{j} }$, respectively.
$V_{\pm (\alpha)}( {\bm r}_j )$  creates (annihilates) a 2$p$ core hole. 
The momentum transfer is defined as $\bm{q}=\bm{k}_{\rm in}-\bm{k}_{\rm out}$. The $R_{\bm{q}, \omega_{\rm in}}$ describes the effective excitation amplitude of a magnon via a local spin flip induced by two x-ray excitations from the altermagnetic ground state.
Then, the propagation of the spin excitation in the lattice is described by $(\omega_{\rm loss} + E_i - H + i\delta)^{-1}$. Note that the propagator does not contain a 2$p$ core hole. We simulate the propagator by replacing $H$ with the linearized spin-wave Hamiltonian for the spin model $\mathcal{H}$ in Sec.~III.

Given that the Mn 2$p$ core hole created by the x-rays is highly localized at the excited site and that MnTe is a wide-gap Mott-Hubbard insulator, the $R(\bm{q}, \omega_{\rm in})$ operator can be represented by a Mn$^{2+}$ atomic model~\cite{Wang.R_etal.Phys.-Rev.-B2018}. We employ the same Mn\( ^{2+} \) atomic model introduced in Ref.~\cite{Hariki.A_etal.Phys.-Rev.-Lett.2024},
which successfully reproduced the Mn $L_{2,3}$-edge x-ray absorption and XMCD spectrum of MnTe.
The atomic model includes the 3$d$--3$d$ and 2$p$--3$d$ Coulomb multiplet interactions, spin-orbit interactions in the 3$d$ and 2$p$ shells, and crystal-field splitting on the Mn ion, see Ref.~\cite{Hariki.A_etal.Phys.-Rev.-Lett.2024} for the parameter values for these terms.  
Note that the unit cell of MnTe contains two Mn sites related by a screw operation. Following Ref.~\cite{Hariki.A_etal.Phys.-Rev.-Lett.2024}, we treat the two Mn sites explicitly, and the interference between the two Mn sites in the RIXS cross-section is taken into account in the approach (i). 

The transition operators $V$ encode the experimental geometry and photon polarization effects~\cite{Hariki.A_etal.Phys.-Rev.-B2020}. We calculate spectra for two independent polarizations in $V_\alpha$ and take the average of the two, as the polarization of the emitted x-rays is not resolved in the present RIXS experiment. The same treatment for the emitted x-rays was applied in the approach (ii) below.

\section{\NoCaseChange{LDA+DMFT RIXS simulation}}


Next, we summarize the approach (ii) that employs the LDA+DMFT method. We use the same implementation as in Refs.~\cite{Hariki.A_etal.Phys.-Rev.-Lett.2018,Hariki.A_etal.Phys.-Rev.-B2020,Li.J_etal.Phys.-Rev.-X2023}. First, we performed a standard LDA+DMFT calculation for MnTe as described in Ref.~\cite{Hariki.A_etal.Phys.-Rev.-Lett.2024}, obtaining the valence electronic structure with the altermagnetic order. Then, we calculated the Mn $L$-edge XMCD and RIXS intensities using the Anderson impurity model (AIM) with the LDA+DMFT spin-polarized hybridization densities $\Delta(\varepsilon)$ representing the Weiss field in the altermagnetically ordered state. Practically, the DMFT hybridization densities $\Delta(\omega)$ were represented by 25 discrete bath levels (per spin and orbital) when computing the RIXS intensities from the AIM.  We used the same parameters for the 3$d$-3$d$ and 2$p$-3$d$ interactions as in the previous study on MnTe~\cite{Hariki.A_etal.Phys.-Rev.-B2020}. Specifically, a Hubbard $U = 5.0$~eV and Hund's coupling $J = 0.86$~eV were used.  
While the AIM description lacks the momentum dependence in the simulated RIXS spectra, it captures the complex evolution of the intermediate states and provides access to various types of excitations (Fig.~\ref{hvmap}), including $dd$ excitations, charge-transfer excitations, and magnetic excitations that are not limited to single-magnon processes but also include two-magnon excitations in the simulated spectra. Thus, the simulated spectrum offers complementary information to the approach (i) which simulates single-magnon processes only, without accounting for the intermediate-state evolution beyond the excited atom.

As described in the main text, the RIXS-CD depends on the domain. Figure~\ref{mcd_phidep}(a) shows the azimuthal angle $\phi$ dependence of the RIXS-CD for the three domains related by the $C_3$ symmetry, computed using the LDA+DMFT AIM approach. The calculated RIXS-CD follows the symmetry properties introduced in the main text. As noted above, MnTe consists of two Mn sites related by a screw operation. Consequently, for general $\phi$ angles, the RIXS-CD contributions from the two Mn sites are different, as illustrated in Fig.~\ref{mcd_phidep}(b).

\bibliography{MnTe}